\documentclass{sig-alternate}

\begin{document}
%
\conferenceinfo{IWCMC'06,} {July 3--6, 2006, Vancouver, British
Columbia, Canada.}
\CopyrightYear{2006}
\crdata{1-59593-306-9/06/0007}


\title{Energy-Efficient Power and Rate Control
 with QoS Constraints: A Game-Theoretic Approach\titlenote{This research was supported by the National Science Foundation
under Grant ANI-03-38807.}}
%
%

\numberofauthors{2}
%

\author{
%
\alignauthor Farhad Meshkati, H. Vincent Poor, Stuart C.
Schwartz\\
       \affaddr{Department of Electrical Engineering}\\
       \affaddr{Princeton University}\\
       \affaddr{Princeton, NJ 08544 USA}\\
       \email{\{meshkati,poor,stuart\}@princeton.edu}
\alignauthor Radu V. Balan\\
       \affaddr{Siemens Corporate Research}\\
       \affaddr{755 College Road East}\\
       \affaddr{Princeton, NJ 08540 USA}\\
       \email{radu.balan@siemens.com}
}

\maketitle

\begin{abstract}
A game-theoretic model is proposed to study the cross-layer problem
of joint power and rate control with quality of service (QoS)
constraints in multiple-access networks. In the proposed game, each
user seeks to choose its transmit power and rate in a distributed
manner in order to maximize its own utility and at the same time
satisfy its QoS requirements. The user's QoS constraints are
specified in terms of the average source rate and average delay. The
utility function considered here measures energy efficiency and the
delay includes both transmission and queueing delays. The Nash
equilibrium solution for the proposed non-cooperative game is
derived and a closed-form expression for the utility achieved at
equilibrium is obtained. It is shown that the QoS requirements of a
user translate into a ``size" for the user which is an indication of
the amount of network resources consumed by the user. Using this
framework, the tradeoffs among throughput, delay, network capacity
and energy efficiency are also studied.
\end{abstract}




\section{Introduction}\label{introduction}

Future wireless networks are expected to support a variety of
services with diverse quality of service (QoS) requirements. Because
of the hostile characteristics of wireless channels and scarcity of
radio resources such as energy and bandwidth, efficient resource
allocation schemes are necessary for design of high-performance
wireless networks. The objective is to use the radio resources as
efficiently as possible and at the same time satisfy the QoS
requirements of the users which can be expressed in terms of
constraints on rate, delay or fidelity.

Since in most practical scenarios, the users' terminals are
battery-powered, energy efficient resource allocation is crucial to
prolonging the battery life of the terminals. In this work, we study
the cross-layer problem of QoS-constrained joint power and rate
control in wireless networks using a game-theoretic framework. We
consider a multiple-access network and propose a non-cooperative
game in which each user seeks to choose its transmit power and rate
in such a way as to maximize its energy-efficiency (measured in bits
per joule) and at the same time satisfy its QoS requirements. The
QoS constraints are in terms of the average source rate and average
total delay (transmission plus queueing delay).  We derive the Nash
equilibrium solution for the proposed game and use this framework to
study tradeoffs among throughput, delay, network capacity and energy
efficiency. Network capacity here refers to the maximum number of
users that can be accommodated by the network.

Joint power and rate control with QoS constraints have been studied
extensively for multiple-access networks (see for example
\cite{Honig96} and \cite{Oh99}). In \cite{Honig96}, the authors
study joint power and rate control under bit-error rate (BER) and
average delay constraints. \cite{Oh99} considers the problem of
globally optimizing the transmit power and rate to maximize
throughput of non-real-time users and protect the QoS of real-time
users. Neither work takes into account energy-efficiency. Recently
tradeoffs between energy efficiency and delay have gained more
attention. The tradeoffs in the single-user case are studied in
\cite{Collins99}--\cite{Fu03}. The multiuser problem in turn is
considered in \cite{Uysal02} and \cite{Coleman04}. In
\cite{Uysal02}, the authors present a centralized scheduling scheme
to transmit the arriving packets within a specific time interval
such that the total energy consumed is minimized whereas in
\cite{Coleman04}, a distributed ALOHA-type scheme is proposed for
achieving energy-delay tradeoffs.  Joint power and rate control for
maximizing goodput in delay-constrained networks is studied in
\cite{Ahmed04}.

This work is the first  study of QoS-constrained power and rate
control in multiple-access networks using a game-theoretic
framework. In our proposed game-theoretic model, users choose their
transmit powers and rates in a \emph{competitive} and
\emph{distributed} manner in order to maximize their energy
efficiency and at the same time satisfy their QoS requirements.
Using this framework, we also analyze the tradeoffs among
throughput, delay, network capacity and energy efficiency. Even
though a centralized approach is possible due to presence of an
access point (AP), a distributed mechanism is more attractive due to
its scalability and low complexity. It should be noted that
game-theoretic approaches to power control have previously been
studied in various work (see for example
\cite{GoodmanMandayam00}--\cite{MeshkatiISIT}). However,
\cite{GoodmanMandayam00}--\cite{MeshkatiJSAC} do not take into
account the effect of delay, and \cite{MeshkatiISIT} only considers
transmission delay and does not perform any rate control.

The remainder of this paper is organized as follows. In
Section~\ref{system model}, we describe the system model. The
proposed joint power and rate control game is discussed in
Section~\ref{PRCG} and its Nash equilibrium solution is derived in
Section~\ref{NE}. We then describe an admission control scheme in
Section~\ref{admission control}. Tradeoffs among throughput, delay,
network capacity and energy efficiency are studied in
Section~\ref{numerical results} using numerical results. Finally, we
give conclusions in Section~\ref{conclusions}.

\section{System Model}\label{system model}

We consider a direct-sequence code-division multiple-access
(DS-CDMA) network and propose a non-cooperative (distributed) game
in which each user seeks to choose its transmit power and rate to
maximize its energy efficiency (measured in bits per joule) while
satisfying its QoS requirements. We specify the QoS constraints of
user $k$ by $(r_k,D_k)$ where $r_k$ is the average source rate and
$D_k$ is the upper bound on average delay. The delay includes both
queueing and transmission delays. The incoming traffic is assumed to
have a Poisson distribution with parameter $\lambda_k$ which
represents the average packet arrival rate with each packet
consisting of $M$ bits. The source rate (in bit per second) is hence
given by $r_k= M \lambda_k$ .

The user transmits the arriving packets at a rate $R_k$ (bps) and
with a transmit power equal to $p_k$ Watts. We consider an
automatic-repeat-request (ARQ) mechanism in which the user keeps
retransmitting a packet until the packet is received at the access
point without any errors. The incoming packets are assumed to be
stored in a queue and transmitted in a first-in-first-out (FIFO)
fashion. The packet transmission time for user $k$ is defined as
\begin{equation}\label{eq1}
    \tau_k = \frac{M}{R_k} + \epsilon_k \simeq \frac{M}{R_k} ,
\end{equation}
where $\epsilon_k$ represents the time taken for the user to receive
an ACK/NACK from the access point. We assume $\epsilon_k$ is
negligible compared to $\frac{M}{R_k}$. The packet success
probability (per transmission) is represented by $f(\gamma_k)$ where
$\gamma_k$ is the received signal-to-interference-plus-noise ratio
(SIR) for user~$k$. The retransmissions are assumed to be
independent. The packet success rate, $f(\gamma)$, is assumed to be
increasing and S-shaped (sigmoidal) with $f(0)=0$ and $f(\infty)=1$.
This is a valid assumption for many practical scenarios as long as
the packet size is reasonably large (e.g., $M=100$ bits).

\begin{figure}
\centering
\includegraphics[width=3.1in]{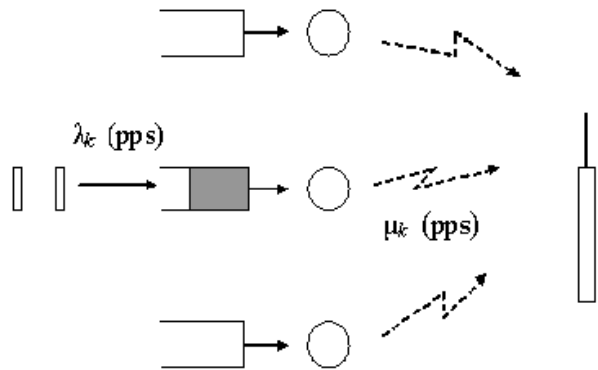}
\caption{System model based on an M/G/1 queue.} \label{fig1-sys}
\end{figure}

We can represent the combination of user $k$'s queue and wireless
link as an M/G/1 queue, as shown in Figure~\ref{fig1-sys}, where the
traffic is Poisson with parameter $\lambda_k$ (in packets per
second) and the service time, $S_k$, has the following probability
mass function (PMF):
\begin{equation}\label{eq2}
    \textrm{Pr}\{S_k=m\tau_k\}= f(\gamma_k)
    \left(1-f(\gamma_k)\right)^{m-1}  \ \ \ \textrm{for} \ m=1, 2,
    \cdots
\end{equation}
As a result, the service rate, $\mu_k$, is given by
\begin{equation}\label{eq4}
    \mu_k=\frac{1}{\mathbb{E}\{S_k\}}= \frac{f(\gamma_k)}{\tau_k}
    ,
\end{equation}
and the load factor $\rho_k=\frac{\lambda_k}{\mu_k}=\frac{\lambda_k
\tau_k}{f(\gamma_k)}$.

To keep the queue of user $k$ stable, we must have $\rho_k<1$ or
$f(\gamma_k)>\lambda_k \tau_k$. Now, let $W_k$ be a random variable
representing the total packet delay for user $k$. This delay
includes the time the packet spends in the queue as well as the
service time. It is known that for an M/G/1 queue the average wait
time (including the queueing and service time) is given by
\begin{equation}\label{eq5}
    \bar{W}_k=\frac{\bar{L}_k}{\lambda_k} ,
\end{equation}
where $\bar{L}_k= \rho_k + \frac{ \rho_k^2 + \lambda_k^2
\sigma_{S_k}^2}{2(1-\rho_k)}$ with $\sigma_{S_k}^2$ being the
variance of the service time \cite{GrossBook85}. Therefore, the
average packet delay for user $k$ is given by
\begin{equation}\label{eq6}
    \bar{W}_k = \tau_k \left( \frac{1-\frac{\lambda_k
    \tau_k}{2}}{f(\gamma_k)-\lambda_k \tau_k} \right) \ \ \
    \textrm{with} \ f(\gamma_k)>\lambda_k \tau_k .
\end{equation}
We require the average delay for user $k$'s packets to be less than
or equal to $D_k$. This translates to
\begin{equation}\label{eq6b}
    \bar{W}_k \leq D_k
\end{equation}
or
\begin{equation}\label{eq7}
    f(\gamma_k) \geq \lambda_k \tau_k + \frac{\tau_k}{D_k} -
    \frac{\lambda_k \tau_k^2}{2D_k} .
\end{equation}
However, since $0 \leq f(\gamma_k) < 1$, \eqref{eq7} is possible
only if\footnote{Note that $f(\gamma)=1$ requires an infinite SIR
which is not practical.}
\begin{equation}\label{eq8}
    0 \leq  \lambda_k \tau_k + \frac{\tau_k}{D_k} -
    \frac{\lambda_k \tau_k^2}{2D_k} < 1.
\end{equation}
This means that $\lambda_k$ and $D_k$ are feasible only if they
satisfy \eqref{eq8}. Note that the upper bound on the average delay
(i.e., $D_k$) cannot be smaller than transmission time $\tau_k$.
This automatically implies that $\lambda_k \tau_k +
\frac{\tau_k}{D_k} -\frac{\lambda_k \tau_k^2}{2D_k}>0$.

Let us define $\eta_k=\lambda_k \tau_k + \frac{\tau_k}{D_k}
-\frac{\lambda_k \tau_k^2}{2D_k}$. Then, \eqref{eq7} is equivalent
to the condition $\gamma\geq \hat{\gamma}_k$ where
\begin{equation}\label{eq9}
    \hat{\gamma}_k=f^{-1}(\eta_k)
\end{equation}
with $\eta_k <1$ and $R_k \geq M/D_k$. This means that the delay
constraint in \eqref{eq6b} translates into a lower bound on the
output SIR. It should be noted that the delay QoS constraint
considered here is in terms of average delay. An average-delay
constraint may not be sufficient for applications with hard delay
requirements (see \cite{MeshkatiDelayTCOM} for a more detailed
discussion on the delay distribution).

\section{The Joint Power and Rate Control Game} \label{PRCG}

Consider the non-cooperative joint power and rate control game
(PRCG) $G=[\mathcal{K}, \{A_k\}, \{u_k\}]$ where
$\mathcal{K}=\{1,2,\cdots,K\}$ is the set of users,
$A_k=[0,P_{max}]\times[0,B]$ is the strategy set for user $k$ with a
strategy corresponding to a choice of transmit power and transmit
rate, and $u_k$ is the utility function for user $k$. Here,
$P_{max}$ and $B$ are the maximum transmit power and the system
bandwidth, respectively. For the sake of simplicity, throughout this
paper, we assume $P_{max}$ is large. Each user chooses its transmit
power and rate in order to maximize its own utility while satisfying
its QoS requirements. The utility function for a user is defined as
the ratio of the user's goodput to its transmit power, i.e.,
\begin{equation}\label{eq10}
    u_k=\frac{T_k}{p_k} ,
\end{equation}
where the goodput $T_k$ is the number of bits that are transmitted
successfully per second and is given by
\begin{equation}\label{eq11}
    T_k= R_k f(\gamma_k) .
\end{equation}
Therefore, the utility function for user $k$ is given by
\begin{equation}\label{eq12}
    u_k=R_k \frac{f(\gamma_k)}{p_k} .
\end{equation}
This utility function has units of bits per joule and is
particularly suitable for wireless networks where energy efficiency
is important.

Fixing other users' transmit powers and rates, the
utility-maximizing strategy for user $k$ is given by the solution of
the following constrained maximization:
\begin{equation}\label{eq12b}
    \max_{p_k, R_k} \ u_k \ \ \ \textrm{s.t.} \ \ \ \bar{W}_k \leq D_k \ ,
\end{equation}
or equivalently
\begin{equation}\label{eq13}
   \max_{p_k, R_k} \ u_k \ \ \ \ \textrm{s.t.} \ \
   \gamma_k>\hat{\gamma}_k \ \ \textrm{and} \ \ 0 \leq \eta_k <1
\end{equation}
where $\hat{\gamma}_k=f^{-1}(\eta_k)$ and
\begin{equation}\label{eq13b}
    \eta_k=\frac{M \lambda_k}{R_k} + \frac{M}{D_k R_k} -\frac{M^2 \lambda_k}{2D_k
    R_k^2} \ .
\end{equation}
Note that for a matched filter receiver and with random spreading
sequences, the received SIR is approximately given by
\begin{equation}\label{eq14}
    \gamma_k= \left(\frac{B}{R_k}\right) \frac{p_k h_k}{\sigma^2 + \sum_{j\neq
    k} p_j h_j} ,
\end{equation}
where $h_k$ is the channel gain for user $k$ and $\sigma^2$ is the
noise power in the bandwidth $B$.

Let us first look at the maximization in \eqref{eq13} without any
constraints, i.e.,
\begin{equation}\label{eq15}
    \max_{p_k, R_k} u_k \ \ \equiv \ \max_{p_k, R_k}  R_k \frac{f(\gamma_k)}{p_k}  .
\end{equation}

\newtheorem{theorem}{Theorem}
\begin{theorem}
The unconstrained utility maximization in \eqref{eq15} has an
infinite number of solutions. More specifically, any combination of
$p_k$ and $R_k$ that achieves an output SIR equal to $\gamma^*$, the
solution to $f(\gamma)=\gamma f'(\gamma)$, maximizes $u_k$.
\end{theorem}
\begin{proof}
Let $\tilde{p}_k$ and $\tilde{R}_k$ be any power-rate combination
such that
$$\left(\frac{B}{\tilde{R}_k}\right) \frac{\tilde{p}_k h_k}{\sigma^2 + \sum_{j\neq k} p_j
h_j}=\tilde{\gamma}.$$ Then, we have
\begin{equation}\label{eq15d}
    \tilde{u}_k= \tilde{R}_k \frac{f(\tilde{\gamma})}{\tilde{p}_k} = \frac{B
    \hat{h}_k}{\tilde{\gamma}} f(\tilde{\gamma}) ,
\end{equation}
where
\begin{equation}\label{eq15c}
    \hat{h}_k = \frac{h_k}{\sigma^2 + \sum_{j\neq
    k} p_j h_j} \ .
\end{equation}
This means that when other users' powers and rates are fixed (i.e.,
fixed $\hat{h}_k$), user $k$'s utility depends only on
$\tilde{\gamma}$ and is independent of the specific values of $p_k$
and $R_k$. In addition, by taking the derivative of
$\frac{f(\gamma)}{\gamma}$ with respect to $\gamma$ and equating it
to zero, it can be shown that $\frac{f(\gamma)}{\gamma}$  is
maximized when $\gamma=\gamma^*$, the (unique) positive solution of
$f(\gamma)=\gamma f'(\gamma)$. Therefore, $u_k$ is maximized for any
combination of $p_k$ and $R_k$ for which $\gamma_k=\gamma^*$. This
means that there are infinitely many solutions for the unconstrained
maximization in \eqref{eq15}.
\end{proof}

The second constraint in \eqref{eq13} can equivalently be expressed
as
\begin{equation}\label{eq18}
    R_k > \left(\frac{M}{D_k}\right) \frac{1+D_k \lambda_k
    +\sqrt{1+ D_k^2 \lambda_k^2}}{2} .
\end{equation}
Therefore, the maximization in \eqref{eq13} is equivalent to
\begin{eqnarray}\label{eq19}
    \max_{p_k, R_k}  && R_k \frac{f(\gamma_k)}{p_k}
    \nonumber\\  \textrm{s.t.} \ \ \gamma_k &>& \hat{\gamma}_k \nonumber\\
     \textrm{and} \ \ R_k &>& \left(\frac{M}{D_k}\right) \frac{1+D_k \lambda_k
    +\sqrt{1+ D_k^2 \lambda_k^2}}{2} \ .
\end{eqnarray}

Let us define $$\Omega_k^{\infty}=\left(\frac{M}{D_k}\right)
\frac{1+D_k\lambda_k +\sqrt{1+ D_k^2 \lambda_k^2}}{2}.$$ Note that
for $R_k=\Omega_k^{\infty}$, we have $\eta_k=1$ and hence
$\hat{\gamma}_k=\infty$. Also, define $\Omega_k^*$ as the rate for
which $\hat{\gamma}_k=\gamma^*$, i.e.,
\begin{equation} \label{eq20}
\Omega_k^*=\left(\frac{M}{D_k}\right) \frac{1+D_k\lambda_k +\sqrt{1+
D_k^2 \lambda_k^2 +2(1-f^*)D_k \lambda_k}}{2f^*}
\end{equation}
where $f^*=f(\gamma^*)$. It is easy to show that $\hat{\gamma}_k >
\gamma^*$ for all $\Omega_k^{\infty}< R_k <\Omega_k^*$.  This means
that based on \eqref{eq15d}, user $k$ has no incentive to transmit
at a rate smaller than $\Omega_k^*$. Furthermore, based on
Theorem~1, any combination of $p_k$ and $R_k \geq \Omega_k^*$ that
results in an output SIR equal to $\gamma^*$ is a solution to the
constrained maximization in \eqref{eq13}. Note that when $R_k =
\Omega_k^*$ and $\gamma_k=\gamma^*$, we have $\bar{W}_k=D_k$.

\section{Nash Equilibrium for the PRCG} \label{NE}

For a non-cooperative game, a Nash equilibrium is defined as a set
of strategies for which no user can unilaterally improve its own
utility \cite{FudenbergTiroleBook91}. We saw in Section~\ref{PRCG}
that for our proposed non-cooperative game, each user has infinitely
many strategies that maximize the user's utility. In particular, any
combination of $p_k$ and $R_k$ for which $\gamma_k=\gamma^*$ and
$R_k\geq\Omega_k^*$ is a best-response strategy.

\begin{theorem}
If  $\sum_{k=1}^K \frac{1}{1+\frac{B}{\Omega_k^* \gamma^*}} < 1$,
then the PRCG has at least one Nash equilibrium given by $(p_k^*,
\Omega_k^*)$,  for $k=1, \cdots , K$,  where
$p_k^*=\frac{\sigma^2}{h_k}\left(\frac{\frac{1}{1+\frac{B}{\Omega_k^*\gamma^*}}}{1-\sum_{j=1}^K
\frac{1}{1+\frac{B}{\Omega_j^*\gamma^*}}}\right)$ and $\Omega_k^*$
is given by \eqref{eq20}. Furthermore, when there are more than one
Nash equilibrium, $(p_k^*, \Omega_k^*)$ is the Pareto-dominant
equilibrium.
\end{theorem}
\begin{proof}
If $\sum_{j=1}^K \frac{1}{1+\frac{B}{\Omega_j^* \gamma^*}} < 1$ then
$p_k^*$ is positive and finite. Now, if we let $p_k=p_k^*$ and
$R_k=\Omega_k^*$, then the output SIR for all the users will be
equal to $\gamma^*$ which means every user is using its
best-response strategy. Therefore, $(p_k^*, \Omega_k^*)$ for
$k=1,\cdots,K$ is a Nash equilibrium.

More generally, if we let $R_k= \tilde{R}_k \geq \Omega_k^*$ and
provided that $\sum_{j=1}^K \frac{1}{1+\frac{B}{\tilde{R}_j
\gamma^*}} < 1$, then $(\tilde{p}_k, \tilde{R}_k)$ is a Nash
equilibrium where
$\tilde{p}_k=\frac{\sigma^2}{h_k}\left(\frac{\frac{1}{1+\frac{B}{\tilde{R}_k
\gamma^*}}}{1-\sum_{j=1}^K \frac{1}{1+\frac{B}{\tilde{R}_j
\gamma^*}}}\right)$. Hence, based on \eqref{eq12}, at Nash
equilibrium, the utility of user $k$ is given by
\begin{equation}\label{eq20b}
    u_k = \frac{B f(\gamma^*) h_k}{\sigma^2 \gamma^*}\left(1-
\frac{\sum_{j \neq k} \frac{1}{1+\frac{B}{\tilde{R}_j
\gamma^*}}}{1-\frac{1}{1+\frac{B}{\tilde{R}_k \gamma^*}}}\right) \ .
\end{equation}
Therefore, the Nash equilibrium with the smallest $\tilde{R}_k$
achieves the largest utility. A higher transmission rate for a user
requires a larger transmit power by that user to achieve $\gamma^*$.
This not only reduces the user's utility but also causes more
interference for other users in the network and forces them to raise
their transmit powers as well which will result in a reduction in
their utilities. This means that the Nash equilibrium with
$R_k=\Omega_k^*$ and $p_k^*$ for $k=1,\cdots,K$ is the
Pareto-dominant equilibrium.
\end{proof}

Based on the feasibility condition given by Theorem~2, let us define
the ``size" of user $k$ as
\begin{equation}\label{eq24}
    \Phi_k^* = \frac{1}{1+\frac{B}{\Omega_k^* \gamma^*}} \ .
\end{equation}
Therefore, the feasibility condition of Theorem~2 can be written as
\begin{equation}\label{eq25}
    \sum_{k=1}^K \Phi_k^* < 1 .
\end{equation}
Note that the QoS requirements of user $k$ (i.e., its source rate
$r_k$ and delay constraint $D_k$) uniquely determine $\Omega_k^*$
through \eqref{eq20} and, in turn, determine the size of the user
(i.e., $\Phi_k^*$) through \eqref{eq24}. The size of a user is
basically an indication of the amount of network resources consumed
by that user. A larger source rate or a tighter delay constraint for
a user increases the size of the user. The network can accommodate a
set of users if and only if their total size is less than 1. In
Section~\ref{numerical results}, we use this framework to study the
tradeoffs among throughput, delay, network capacity and energy
efficiency.

\section{Admission Control}\label{admission control}

In Section~\ref{NE}, we defined the ``size" of a user based on its
QoS requirements. Before joining the network, each user calculates
its size using \eqref{eq24} and announces it to the access point.
According to \eqref{eq25}, the access point admits those users whose
total size is less than 1. While the goal of each user is to
maximize its own energy efficiency, a more sophisticated admission
control can be performed to maximize the total network utility. In
other words, out of the $K$ users, the access point can choose those
users for which the total network utility is the largest, i.e.,
\begin{equation}\label{eq26}
    \max_{{\mathcal{L}}\subset \{1,\cdots,K\}} \ \sum_{\ell \in {\mathcal{L}}} u_{\ell}
\end{equation}
under the constraint that $\sum_{\ell \in {\mathcal{L}}}
\Phi_{\ell}^* <1$.

Based on \eqref{eq20b}, the utility of user $\ell$ at the
Pareto-dominant Nash equilibrium is given by
\begin{equation}\label{eq28}
u_{\ell}=\left(\frac{B h_{\ell} f(\gamma^*)}{\sigma^2
\gamma^*}\right)\frac{1-\sum_{i \in
    {\mathcal{L}}}\Phi_i^*}{1-\Phi_{\ell}^*} \ .
\end{equation}
As a result, \eqref{eq26} becomes
\begin{equation}\label{eq29}
      \max_{{\mathcal{L}}\subset \{1,\cdots,K\}} \left(1-\sum_{i \in
    {\mathcal{L}}}\Phi_i^*\right)  \sum_{\ell \in
     {\mathcal{L}}}  \frac{h_{\ell}}{1-\Phi_{\ell}^*}\
\end{equation}
under the constraint that $\sum_{\ell \in {\mathcal{L}}}
\Phi_{\ell}^* <1$.

In general, obtaining a closed-form solution for \eqref{eq29} is
difficult since it depends on the channel gains and sizes of the
users. Instead, in order to gain some insight, let us consider the
special case in which all users are at the same distance from the
access point. We first consider the scenario in which the users have
identical QoS requirements (i.e.,
$\Phi_1^*=\cdots=\Phi_K^*=\Phi^*$). If we replace $\sum_{\ell=1}^L
h_\ell$ by $L \mathbb{E}\{h\}$, then \eqref{eq29} becomes
\begin{equation}\label{eq30}
    \max_L \frac{\mathbb{E}\{h\} ( L-L^2\Phi^* )}{1-\Phi^*} .
\end{equation}
Therefore, the optimal number of users for maximizing the total
utility in the network is $L=\left[\frac{1}{2\Phi^*}\right]$ where
$[x]$ represents the integer nearest to $x$.

Now consider another scenario in which there are $C$ classes of
users. The users in class $c$ are assumed to all have the same QoS
requirements and hence the same size, $\Phi^{*(c)}$. Since we are
assuming that all the users are at the same distance from the access
point, they all have the same average channel gain. Now, if the
access point admits $L^{(c)}$ users from class $c$ then the total
utility is given by
$$u_{T}=\left(\frac{B \mathbb{E}\{h\} f(\gamma^*)}{\sigma^2 \gamma^*}\right)
\left(1-\sum_{c=1}^C L^{(c)}\Phi^{*(c)}\right)\left(\sum_{c=1}^C
\frac{L^{(c)}}{1-\Phi^{*(c)}}\right)$$ provided that $\sum_{c=1}^C
L^{(c)}\Phi^{*(c)}<1$. Without loss of generality, let us assume
that $\Phi^{*(1)}<\Phi^{*(2)}< \cdots <\Phi^{*(C)}$. It can be shown
that $u_T$ is maximized when
$L^{(1)}=\left[\frac{1}{2\Phi^{*(1)}}\right]$ with $L^{(c)}=0$ for
$c=2, 3, \cdots, C$. This is because adding a user from class~1 is
always more beneficial in terms of increasing the total utility than
adding a user from any other class. Therefore, in order to maximize
the total utility in the network, the access point must admit only
users from the class with the smallest size. While this solution
maximizes the total network utility, it is not fair. A more
sophisticated admission control mechanism can be used to improve the
fairness. In the next section, we demonstrate the loss in network
energy efficiency if a suboptimal admission control strategy is
used.

\section{Numerical Results}\label{numerical results}

Let us consider the uplink of a DS-CDMA system with a total
bandwidth of 5MHz (i.e., $B=5$MHz). Each user in the network has a
set of QoS requirements expressed as $(r_k, D_k)$ where $r_k$ is the
source rate and $D_k$ is the delay requirement (upper bound on the
average total delay) for user~$k$. As explained in Section~\ref{NE},
the QoS parameters of a user define a ``size" for that user, denoted
by $\Phi_k^*$ given by \eqref{eq24}. Before a user starts
transmitting, it must announce its size to the access point. Based
on the particular admission policy, the access point decides whether
or not to admit the user. For the cases where there are multiple
equilibria, we assume that the admitted users agree to choose the
transmit powers and rates that correspond to the Pareto-dominant
Nash equilibrium.  This is an important assumption that requires its
own study.

\begin{figure}
\begin{center}
\leavevmode \hbox{\epsfysize=8cm \epsfxsize=8.5cm
\epsffile{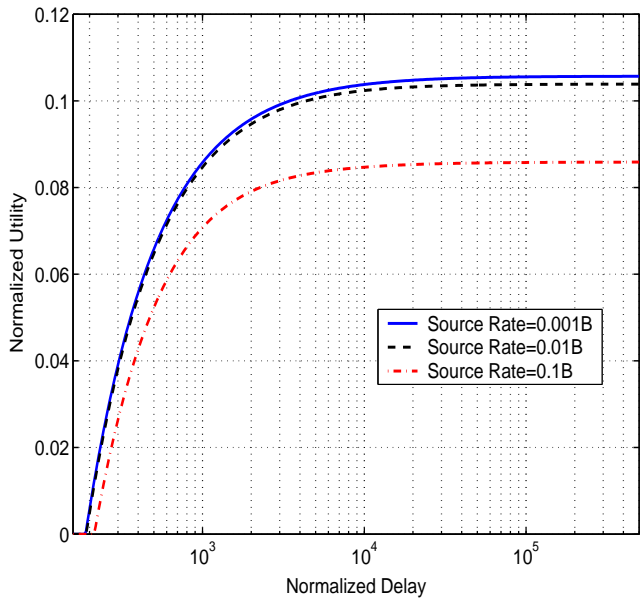}}
\end{center}\vspace{-0.2cm}
\caption{Normalized utility as a function of normalized delay for
different source rates ($B=5$MHz). The combined ``size" of other
users in the network is equal to 0.2.} \label{utilvsdelay}
\end{figure}

Figure~\ref{utilvsdelay} shows the user's utility as a function of
delay for different source rates. The total size of the other users
in the network is assumed to be 0.2. The user's utility is
normalized by $Bh/\sigma^2$, and the delay is normalized by the
inverse of the system bandwidth. As expected, a tighter delay
requirement and/or a higher source rate results in a lower utility
for the user.

\begin{figure}
\begin{center}
\leavevmode \hbox{\epsfysize=9cm \epsfxsize=8.5cm
\epsffile{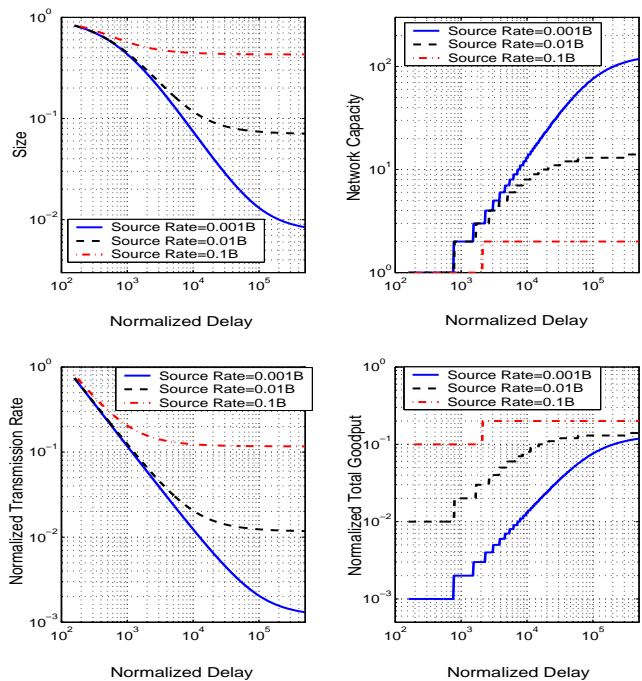}}
\caption{User size, network capacity, normalized transmission rate,
and normalized total goodput as a function of normalized delay for
different source rates ($B=5$MHz).} \label{delayQoS}
\end{center}
\end{figure}

Figure~\ref{delayQoS} shows the user size, network capacity,
transmission rate, and total goodput as a function of normalized
delay for different source rates. The network capacity refers to the
maximum number of users that can be admitted into the network
assuming that all the users have the same QoS requirements (i.e.,
the same size). The transmission rate and goodput are normalized by
the system bandwidth. The total goodput is obtained by multiplying
the source rate by the total number of users. As the QoS
requirements become more stringent (i.e., a higher source rate
and/or a smaller delay), the size of the user increases which means
more network resources are required to accommodate the user. This
results in a reduction in the network capacity. It is also observed
from the figure that when the delay constraint is loose, the total
goodput is almost independent of the source rate. This is because a
lower source rate is compensated by the fact that more users can be
admitted into the network. On the other hand, when the delay
constraint in tight, the total goodput is higher for larger source
rates.
\vspace{0cm}

Now to study admission control, let us consider a network with three
different classes of users/sources:
\begin{enumerate}
  \item Class $A$ users for which $r^{(A)}=5$kbps and $D^{(A)}=10$ms, and hence, $\Phi^{*^{(A)}}=0.0198$.
  \item Class $B$ users for which $r^{(B)}=50$kbps and $D^{(B)}=50$ms, and hence, $\Phi^{*^{(B)}}=0.0718$.
  \item Class $C$ users for which $r^{(C)}=150$kbps and $D^{(C)}=1000$ms, and hence, $\Phi^{*^{(B)}}=0.1848$.
\end{enumerate}
For the purpose of illustration, let us assume that there are a
large number of users in each class and that they all are at the
same distance from the access point (i.e., they all have the same
average channel gain). The access point receives requests from the
users and has to decide which ones to admit in order to maximize the
total utility in the network (see \eqref{eq29}). We know from
Section~\ref{admission control} that since users in class $A$ have
the smallest size, the total utility is maximized if the access
point picks users from class $A$ only with $L^{(A)}=
\left[1/2\Phi^{*^(A)}\right]=25$. However, this solution does not
take into account fairness issues. Instead, we may be more
interested in solutions in which more than one class of users are
admitted. Table~\ref{table1} shows the percentage loss in the total
utility for several choices of $L^{(A)}, L^{(B)}$ and $L^{(C)}$.
\begin{table}
\begin{center} \caption{ Percentage loss in the total network utility for different choices of $L^{(A)},
L^{(B)}$ and $L^{(C)}.$}\label{table1}
\begin{tabular}{|c|c|c|c|}
  \hline
  $L^{(A)}$ & $L^{(B)}$ & $L^{(C)}$ & Loss in total utility \\
  \hline
  \hline
  25 & 0 & 0 & -- \\
   23 & 1 & 0 & 10\% \\
  20 & 0 & 1 & 30\% \\
  18 & 1 & 1 &  38\% \\
  0 & 7 & 0 & 71\% \\
  0 & 0 & 3 & 87\% \\
  \hline
\end{tabular}
\end{center}
\vspace{-0.55cm}
\end{table}

\section{Conclusions}\label{conclusions}

We have studied the cross-layer problem of joint power and rate
control with QoS constraints in wireless networks using a
game-theoretic framework. We have proposed a non-cooperative game in
which users seek to choose their transmit powers and rates in such a
way as to maximize their utilities and at the same time satisfy
their QoS requirements. The utility function considered here
measures the number of reliable bits transmitted per joule of energy
consumed. The QoS requirements for a user consist of the average
source rate and the average delay where the delay includes both
transmission and queueing delays. We have derived the Nash
equilibrium solution for the proposed game and obtained a
closed-form solution for the user's utility at equilibrium. Using
this framework, we have studied the tradeoffs among throughput,
delay, network capacity and energy efficiency, and have shown that
presence of users with stringent QoS requirements results in
significant reductions in network capacity and energy efficiency.


\begin{thebibliography}{10}
\vspace{0.2cm}
\bibitem{Honig96} M.~L. Honig and J.~B. Kim, ``Allocation of
{DS-CDMA} parameters to achieve
  multiple rates and qualities of service,'' {\em Proceedings of the IEEE
  Global Telecommunications Conference (Globecom)}, pp.~1974--1978, London, England,
  Nov.
  1996.

\bibitem{Oh99}
S.-J. Oh and K.~M. Wasserman, ``Adaptive resource allocation in
power
  constrained {CDMA} mobile networks,'' {\em Proceedings of the IEEE Wireless
  Communications and Networking Conference (WCNC)}, pp.~510--514, New Orleans, LA, USA,
  Sep. 1999.

\bibitem{Collins99}
B.~Collins and R.~Cruz, ``Transmission policies for time varying
channels with
  average delay constraints,'' {\em Proceedings of the $37^{th}$ Annual
  Allerton Conference on Communication, Control, and Computing}, Monticello,
  IL, USA, Oct. 1999.

\bibitem{Prabhakar01}
B.~Prabhakar, E.~Uysal-Biyikoglu, and A.~El~Gamal,
``Energy-efficient
  transmission over a wireless link via lazy packet scheduling,'' {\em
  Proceedings of $20^{th}$ Annual Joint Conference of the IEEE Computer and
  Communications Societies (INFOCOM)}, Anchorage, AK, USA, Apr. 2001.

\bibitem{Berry02}
R.~A. Berry and R.~G. Gallager, ``Communication over fading channels
with delay
  constraints,'' {\em IEEE Transactions on Information Theory}, vol.~48,
  pp.~1135--1149, May 2002.

\bibitem{Fu03}
A.~Fu, E.~Modiano, and J.~Tsitsiklis, ``Optimal energy allocation
for
  delay-constrained data transmission over a time-varying channel,'' {\em
  Proceedings of $22^{nd}$ Annual Joint Conference of the IEEE Computer and
  Communications Societies (INFOCOM)}, San Francisco, CA, USA, Mar./Apr. 2003.

\bibitem{Uysal02}
E.~Uysal-Biyikoglu and A.~El~Gamal, ``Energy-efficient packet
transmission over
  multiaccess channel,'' {\em Proceedings of IEEE International Symposium on
  Information Theory (ISIT)}, Lausanne, Switzerland, Jun./Jul. 2002.

\bibitem{Coleman04}
T.~P. Coleman and M.~M\'{e}dard, ``A distributed scheme for
achieving
  energy-delay tradeoffs with multiple service classes over a dynamically
  varying network,'' {\em IEEE Journal on Selected Areas in Communications
  (JSAC)}, vol.~22, pp.~929--941, Jun. 2004.

\bibitem{Ahmed04}
N.~Ahmed, M.~A. Khojestapour, and R.~G. Baraniuk, ``Delay-limited
throughput
  maximization for fading channels using rate and power control,'' {\em
  Proceedings of the IEEE Global Telecommunications Conference (Globecom)},
  pp.~3459--3463, Dallas, TX, USA, Nov./Dec. 2004.

\bibitem{GoodmanMandayam00}
D.~J. Goodman and N.~B. Mandayam, ``Power control for wireless
data,'' {\em
  IEEE Personal Communications}, vol.~7, pp.~48--54, Apr. 2000.

\bibitem{Feng99}
N.~Feng, N.~B. Mandayam, and D.~J. Goodman, ``Joint power and rate
optimization
  for wireless data services based on utility functions,'' {\em Proceedings of
  the $33^{rd}$ Annual Conference on Information Sciences and Systems (CISS)},
  Baltimore, MD, USA, Mar. 1999.

\bibitem{Alpcan01}
T.~Alpcan, T.~Basar, R.~Srikant, and E.~Altman, ``{CDMA} uplink
power control
  as a noncooperative game,'' {\em Proceedings of the $40^{th}$ {IEEE}
  Conference on Decision and Control}, pp.~197--202, Orlando, FL, USA,
  Dec. 2001.

\bibitem{Xiao01}
M.~Xiao, N.~B. Shroff, and E.~K.~P. Chong, ``Utility-based power
control in
  cellular wireless systems,'' {\em Proceedings of the Annual Joint Conference
  of the IEEE Computer and Communications Societies (INFOCOM)}, pp.~412--421,
  Anchorage, AK, USA, Apr. 2001.

\bibitem{Sung02}
C.~W. Sung and W.~S. Wong, ``A noncooperative power control game for
multirate
  {CDMA} data networks,'' {\em IEEE Transactions on Wireless Communications},
  vol.~2, pp.~186--194, Jan. 2003.

\bibitem{MeshkatiTcomm}
F.~Meshkati, H.~V. Poor, S.~C. Schwartz, and N.~B. Mandayam, ``An
  energy-efficient appraoch to power control and receiver design in wireless
  data networks,'' {\em IEEE Transactions on Communications}, vol.~52,
  pp.~1885--1894, Nov. 2005.

\bibitem{MeshkatiJSAC}
F.~Meshkati, M.~Chiang, H.~V. Poor, and S.~C. Schwartz, ``A
game-theoretic
  approach to energy-efficient power control in multi-carrier {CDMA} systems,''
  to appear in the {\it{IEEE Journal on Selected Areas in Communications
  (JSAC): Special Issue on Advances in Multicarrier CDMA}}, 2006.

\bibitem{MeshkatiISIT}
F.~Meshkati, H.~V. Poor, and S.~C. Schwartz, ``A non-cooperative
power control
  game in delay-constrained multiple-access networks,'' {\em Proceedings of the
  IEEE International Symposium on Information Theory (ISIT)}, Adelaide,
  Australia, Sep. 2005.

\bibitem{GrossBook85}
D.~Gross and C.~M. Harris, {\em Fundamentals of Queueing Theory}.
\newblock John Wiley \& Sons, 1985.

\bibitem{MeshkatiDelayTCOM}
F.~Meshkati, H.~V. Poor, S.~C. Schwartz, and R.~V. Balan,
``Energy-efficient
  resource allocation in wireless networks with quality-of-service
  constraints,'' preprint, Princeton University, 2005.

\bibitem{FudenbergTiroleBook91}
D.~Fudenberg and J.~Tirole, {\em Game Theory}.
\newblock MIT Press, Cambridge, MA, 1991.

\end{thebibliography}
\small{{

}}

\end{document}